\documentclass{article}

\usepackage{amsmath, amssymb}
\usepackage{physics}

\usepackage{graphicx}
\usepackage{tikz}
\usetikzlibrary{positioning, arrows.meta}

\usepackage{algorithm}
\usepackage{algpseudocode}

\usepackage{listings}
\usepackage{xcolor}

\usepackage{cite}

\usepackage{authblk}

\usepackage{hyperref}

\title{Low-cost quantum error mitigation via auxiliary qubit return validation}

\author{Gilad Kishony}
\author{Avi Elazari}
\author{Ron Cohen}
\author{Lior Gazit}

\affil{Classiq Technologies, 3 Daniel Frisch Street, Tel Aviv-Yafo, 6473104, Israel}

\date{}

\begin{document}

\maketitle

\begin{abstract}
We introduce a low-overhead, straightforward technique for quantum error mitigation based on post-selection with auxiliary qubit measurements. The method leverages the structural property that, in a fault-free computation, auxiliary qubits are often expected to return to the $\ket{0}$ state after use. By selectively measuring these auxiliary qubits at carefully chosen points in the circuit, we identify and discard erroneous shots, thereby improving result fidelity with minimal hardware cost. To account for circuit-level noise, including measurement errors, we analyze the likelihood that a given measurement outcome indicates a corrupted shot. This analysis is informed by each measurement’s backward lightcone - the operations in the circuit that could affect the outcome. Shots whose auxiliary measurement outcomes imply a likelihood of corruption above a tunable threshold are rejected. Simulations show that this method reduces the false-negative rate by ~10\% while discarding only ~1\% of valid shots. This threshold controls the bias-variance tradeoff inherent to post-selection, allowing the method to be tailored to the specific fidelity and sampling requirements of a given application.

\end{abstract}

\section{Introduction}
Quantum computations are highly susceptible to noise, especially in the near-term Noisy Intermediate-Scale Quantum (NISQ) regime. While full quantum error correction remains currently out of reach, error mitigation techniques extend the size of quantum circuits from which meaningful results can still be extracted \cite{Preskill2018nisq,Temme2017error,Endo2018practical,Cai2022review,Li2017zne,McArdle2019symmetry,BonetMonroig2018symmetry,Huggins2021vdistill,Czarnik2021clifford}. In this work, we introduce a low-cost error mitigation method that uses mid-circuit and final measurements of auxiliary qubits to validate their expected return to the $\ket{0}$ state, followed by post-selection based on these outcomes.

Auxiliary qubits are frequently used in quantum algorithms for entanglement generation or circuit control. In many designs, these auxiliary qubits are expected to return to the $\ket{0}$ state after completing their task. In a noise-free setting, this is guaranteed by design. However, in practice, noise and gate errors may corrupt the auxiliary’s state.

As a motivating example, consider an arithmetic computation compiled into a sequence of functional blocks, each of which allocates auxiliary qubits to store intermediate results and then explicitly uncomputes them once the block is completed. By construction, all auxiliary qubits are expected to return to the $\ket{0}$ state at the end of each block. Any deviation from this behavior, therefore, indicates the occurrence of a circuit-level error, either during the computation itself or during the uncomputation phase. This structural property, which naturally arises in compiler-generated arithmetic circuits, serves as the basis for the error-mitigation strategy studied in this work.

Our idea is to leverage this structural expectation. If we measure an auxiliary at one of these points and observe a $\ket{1}$, this is undoubtedly an indication of the occurrence of an error. We can thus reject the corresponding shot in a classical postprocessing step. However, this raises several questions about measurement errors, error propagation, and runtime efficiency, which we explore in this paper. Although our method does not offer a bias-free limit, its main advantage is that it requires minimal adjustment to execute the quantum circuits. Furthermore, our method can be easily combined with many other methods for error correction or mitigation.

\section{A Simple Likelihood Model for Post-selection}

Suppose measurements are added to verify the return of auxiliary qubits to the $\ket{0}$ state at $k$ locations within a given circuit. Let $\vec{m} = (m_1, \dots, m_k) \in \{0,1\}^k$ be the vector of observed auxiliary outcomes during a single execution of the circuit. In the absence of noise, all auxiliary measurements are expected to yield $m_i = 0$. Deviations from this ideal, i.e., $m_i = 1$ at some location $i$, may result either from gate-induced errors propagating to the auxiliary or from independent measurement noise. Measurement errors can be a dominant noise source in some hardware, so blindly rejecting all shots with imperfect outcomes may be too wasteful. Instead, we now construct a simple probabilistic model to estimate the likelihood that a given shot is uncorrupted, given the observed auxiliary measurement outcomes. For each measurement $m_i$, let $G_i$ denote the number of gates in its backward lightcone, and suppose:

\begin{itemize}
    \item Each gate fails independently with probability $p$
    \item When a gate fails within the lightcone of a measurement $m_i$, the outcome flips with probability $r$
    \item Each measurement independently reports the wrong outcome with probability $q$
\end{itemize}

For each such measurement location $i$, the outcome $m_i = 1$ could be due either to a gate error or to measurement noise. In the low-error regime, we assume these are disjoint alternatives, and approximate the conditional probability that the true (pre-measurement) state of $m_i$ is uncorrupted as
\begin{equation}
    \Pr(\text{no error in lightcone $i$} \mid m_i = 1) \approx \frac{q}{G_i r p+ q}.
\end{equation}

Assuming independence across measurements (this is not the case when lightcones of different measurements intersect), the probability that \emph{none} of the observed $m_i = 1$ outcomes are due to gate errors is given by
\begin{equation}
    \mathcal{P}(\vec{m})\equiv\Pr(\text{no error in any lightcone} \mid \vec{m}) \approx \prod_{i : m_i = 1} \frac{q}{G_i r p + q}.
\end{equation}
Of course, errors occurring outside the union of the light cones of all auxiliary measurements are never detected.

This model provides a simple and interpretable basis for post-selection: one may reject all shots for which this estimated probability falls below a chosen threshold, i.e. $\mathcal{P}(\vec{m})<\mathcal{P}_\text{th}$, trading off between statistical variance and systematic bias. When all lightcones are of roughly the same size $G_i\approx G$, this simply places a threshold on the number of measurement outcomes of $1$ we are willing to retain.

\section{Bias-Variance Trade-off in Post-selection}
\label{sec:bias_variance}

The method presented naturally introduces a tunable bias-variance trade-off: applying a more stringent post-selection threshold, i.e., accepting only shots with a high probability of being uncorrupted, reduces systematic bias at the expense of statistical variance, due to the decreased number of retained shots.

Formally, denote by $\Pr(\text{retain} \mid \text{good})$ the probability of retaining an error-free (uncorrupted) shot and by $\Pr(\text{retain} \mid \text{corrupted})$ the probability of mistakenly retaining a shot corrupted by gate errors within the auxiliary measurement lightcones. Additionally, shots corrupted by errors occurring outside these lightcones remain unaffected by auxiliary validation and thus pass through the post-selection filter with probability $\Pr(\text{retain}\mid\text{good})$.

The overall fraction of retained shots can then be expressed roughly as
\begin{align}
f_{\text{retain}} &= \left(f_{\text{good}} + f_{\text{undetectable}}\right) \cdot \Pr(\text{retain} \mid \text{good}) \nonumber\\
&+ f_{\text{detectable}} \cdot \Pr(\text{retain} \mid \text{corrupted})
\end{align}

where $f_{\text{good}}$, $f_{\text{detectable}}$, and $f_{\text{undetectable}}$ are the fractions of executed shots that are respectively uncorrupted, corrupted by gate errors within auxiliary measurement lightcones, and corrupted by gate errors outside these lightcones.

The bias of the estimator for any observable can be approximated by
\begin{equation}
\text{Bias} = \Delta \cdot \frac{f_{\text{detectable}} \cdot \Pr(\text{retain}\mid \text{corrupted}) + f_{\text{undetectable}} \cdot \Pr(\text{retain}\mid\text{good})}{f_{\text{retain}}},
\end{equation}
\label{eq_var}
where $\Delta$ represents the typical deviation in an observable of interest caused by a corrupted shot.

The overhead of this mitigation method is defined by the loss of samples, specifically the discarded fraction $1-f_{retain}$. This reduction in retained shots is directly responsible for the increased statistical variance modeled in Equation \ref{eq_var}.

Conversely, the variance of the estimator increases as fewer shots are retained, and can be expressed as
\begin{equation}
\text{Variance} = \frac{\sigma^2}{N \cdot f_{\text{retain}}},
\end{equation}
where $\sigma^2$ is the intrinsic variance of the observable and $N$ is the total number of executed shots. The optimal balance between bias and variance depends on the application scenario.
% \begin{itemize}
% \item For applications requiring high-fidelity results, such as variational quantum algorithms, aggressive post-selection may be beneficial to minimize bias, despite the higher variance.
% \item In contrast, for tasks involving statistical sampling or estimation where variance reduction is critical, a less stringent threshold that accepts more shots, even if slightly corrupted, may be preferred.
% \end{itemize}
Our approach allows adaptive selection of the post-selection threshold, enabling practitioners to effectively tailor the bias-variance trade-off to their specific requirements.

From a practical perspective, auxiliary-based validation is most effective in applications that prioritize estimator fidelity over raw sampling throughput, such as variational algorithms,  expectation-value estimation, or scenarios in which only a limited number of high-quality samples are required. In contrast, applications dominated by large-scale statistical sampling may benefit from a more conservative validation threshold in order to avoid excessive variance from shot rejection. The tunable post-selection threshold therefore, allows the method to adapt to a broad range of application regimes.

\section{Selecting Measurement Locations}
\label{sec:measurement_selection}

Incorporating measurements for auxiliary validation involves a careful balance between diagnostic power and practical implementation costs. The diagnostic power is determined primarily by two factors: the size of the backward lightcone associated with the measurement and its overlap with the lightcones of other measurements. Larger, non-overlapping lightcones typically provide better coverage, allowing the detection of a larger fraction of the possible gate errors that may occur.

However, depending on the location within the circuit, each measurement may introduce additional overhead in terms of execution latency and may induce new errors. Therefore, measurements should be selected strategically, prioritizing those that offer high diagnostic power at minimal latency and error insertion costs. For instance, one may choose measurement points at positions where auxiliary qubits are naturally idle in any case, thus minimizing disruption.

In practice, selecting such validation points does not require manual circuit inspection. When circuits are synthesized using a high-level compiler such as Classiq \cite{classiq} that tracks qubit lifetimes and reuse, all points at which auxiliary qubits are expected to be uncomputed and returned to the $\ket{0}$ state can be identified automatically. Furthermore, using Classiq's compiler, one can automatically generate different circuits for the same high-level model, allocating auxiliaries differently. This compiler-level visibility and control transforms auxiliary validation from a circuit-design challenge into a systematic, scalable procedure.

\section{Early Abort Opportunities}
In addition to postselection-based error mitigation, mid-circuit auxiliary validation enables early termination of corrupted executions, providing an independent and potentially significant reduction in quantum processing unit (QPU) runtime.

Early measurement outcomes during circuit execution may be sufficient to indicate a high likelihood of a detected error. Once we observe these outcomes, executing the remainder of the quantum circuit may no longer be beneficial. By promptly aborting execution upon detecting such events, we conserve valuable QPU time.

The effectiveness of early abort strategies depends heavily on hardware constraints and the nature of the computational task. For deep arithmetic circuits with extensive auxiliary reuse, requiring only a small number of high-fidelity samples, such early-abort decisions can prevent the execution of multiple subsequent functional blocks and can substantially reduce total QPU runtime.

\section{Simulation Results}
\label{sec:simulation_results}

To evaluate the practical effectiveness of auxiliary-based validation, we performed
numerical simulations under a realistic gate-level noise model (stochastic gate and
measurement errors) using circuits synthesized automatically by the Classiq platform
\cite{classiq}.

We study an arithmetic computation that decomposes into multiple functional blocks
and makes extensive use of auxiliary qubits. In the synthesized program, each block
is implemented by a forward computation followed by its inverse, so that the
auxiliaries allocated within the block are uncomputed and return to the $\ket{0}$
state before being reused by later blocks. Deviations from this expected return-to-zero
Behavior, therefore, provides a natural indicator of circuit-level errors.

\begin{figure}[t]
    \centering
    \includegraphics[width=1\linewidth]{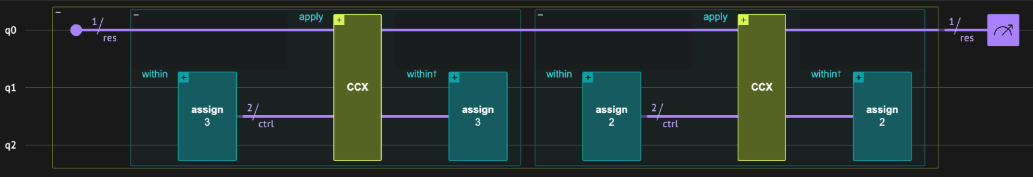}
    \caption{Example quantum circuit synthesized by the compiler, illustrating
    reuse of auxiliary qubits across multiple arithmetic functional blocks. Auxiliary
    qubits are allocated for intermediate computations and uncomputed at the end of
    each block, returning to the $\ket{0}$ state at the indicated reset points.}
    \label{fig:auxiliary_reuse_circuit}
\end{figure}

A representative circuit illustrating auxiliary allocation, uncomputation, and reuse
across blocks is shown in Figure~\ref{fig:auxiliary_reuse_circuit}. A substantially
larger instance of the same computation, exhibiting extensive auxiliary reuse and
circuit depth is available as an interactive visualization in
Ref.~\cite{classiq_large_circuit}.

We compare three auxiliary validation strategies:
\begin{enumerate}
    \item No auxiliary validation.
    \item Validation only at the final circuit measurement.
    \item Validation at every expected auxiliary reset point.
\end{enumerate}

Circuit instances of different widths and depths were generated automatically by
varying auxiliary allocation and reuse strategies within the compiler, allowing us
to probe different regimes for the same logical computation. Because the compiler
tracks qubit lifetimes, and expected auxiliary reset points were identified automatically,
enabling systematic insertion of validation measurements without manual inspection.

\subsection{Arithmetic Model and QMOD Implementation}
\label{subsec:qmod_example}

Classiq’s high-level QMOD abstraction expresses the arithmetic computation compactly. An example is shown in Listing~\ref{lst:qmod_example}, defining a sequence of arithmetic operations (addition, multiplication, and comparison) that introduce intermediate values which are later uncomputed.

\begin{lstlisting}[language=Python, caption={QMOD arithmetic model generating a deep circuit with extensive auxiliary reuse.}, label={lst:qmod_example}]
@qfunc
def main(z: Output[QNum]):
    x = QNum()
    y = QNum()

    x |= 2
    y |= 1

    z |= (2 * x + y + max(3 * y, 2)) > 4
\end{lstlisting}

During synthesis, each arithmetic sub-expression is compiled into a dedicated block that allocates auxiliaries to store intermediate results and then applies its inverse once the contribution has been incorporated. This structure yields circuits with deep execution and nontrivial auxiliary lifetimes, such as the instance in Ref.~\cite{classiq_large_circuit}, and it defines the reset points used for auxiliary validation in the following sections.

\subsection{False Positive vs.\ False Negative Tradeoff}
\label{subsec:fp_fn_tradeoff}

\begin{figure}[t]
    \centering
    \includegraphics[width=0.8\linewidth]{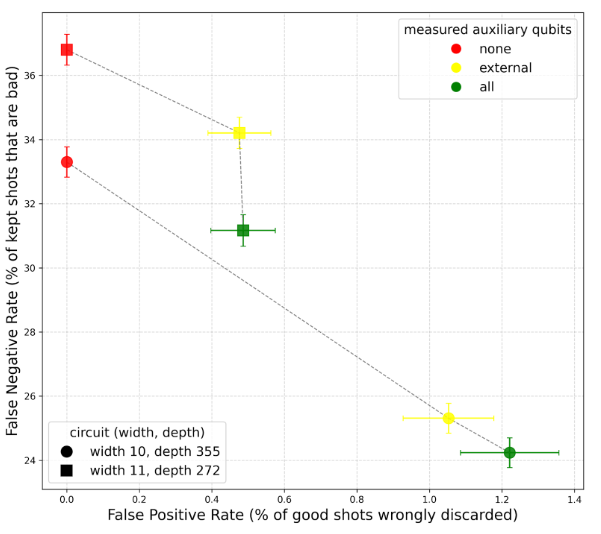}
    \caption{Tradeoff between false positive rate and false negative rate under
    a realistic noise model. Colors indicate the auxiliary validation strategy:
    no validation, validation only at the final measurement, and validation at
    all expected auxiliary reset points. Different markers correspond to
    circuit implementations with varying width and depth.}
    \label{fig:false_positive_negative_tradeoff}
\end{figure}

Figure~\ref{fig:false_positive_negative_tradeoff} shows the tradeoff between false positive rates (valid shots incorrectly discarded) and false negative rates (corrupted shots incorrectly retained) for the three auxiliary validation strategies across two different circuit implementations.

We observe that validating auxiliary qubits at all expected reset points substantially reduces the false-negative rate by approximately 10\% relative to the no-validation baseline, while increasing the false-positive rate by only about 1\%. This ~1\% false-positive rate represents the practical overhead of our method; it is the exact percentage of uncorrupted, valid samples that are lost to post-selection. Validation only at the end of the circuit yields an intermediate improvement, demonstrating that mid-circuit validation provides additional diagnostic power beyond final-state checks alone.

These results empirically confirm the bias--variance tradeoff discussed in Section~\ref{sec:bias_variance}, showing that aggressive validation can significantly suppress systematic bias at a modest cost in statistical variance.

\subsection{Impact of Measurement Location}
\label{subsec:measurement_location}

The comparison between end-only validation and validation at all expected reset
points highlights the importance of measurement placement. Mid-circuit
measurements associated with auxiliary reuse detect errors that would otherwise
propagate undetected to later stages of the computation.

In arithmetic circuits with repeated auxiliary allocation and uncomputation, errors occurring early in the execution may corrupt logical qubits while leaving final auxiliary states apparently valid. Validation at intermediate reset points, therefore, provides increased coverage by associating measurements with smaller, more localized backward lightcones.

This observation supports the heuristic discussed in Section~\ref{sec:measurement_selection}: validation measurements with large backward lightcones and limited overlap provide superior diagnostic coverage.

\section{Discussion}
\label{sec:discussion}

We have introduced a practical, low-overhead method for quantum error mitigation based on auxiliary validation and postselection. 
% By balancing measurement reliability, cost, and lightcone size, our method adapts to a variety of circuit structures and hardware constraints.
Our simulation results demonstrate that auxiliary validation is a practical and effective error mitigation strategy for near-term quantum devices. The method requires minimal circuit modification, integrates naturally with compiler-level qubit reuse analysis, and can be combined with other error mitigation or error-detection techniques. Importantly, the results show that significant reductions in estimator bias can be achieved with only a modest increase in variance, corresponding to an overhead of discarding only about 1\% of valid samples, making the approach particularly appealing.

\bibliographystyle{plain}
\bibliography{references}

\end{document}